 \documentstyle[pre,aps,floats,twocolumn,psfig]{revtex}

\begin{document}

\def\R{\mbox{$I\!\!R$}}

\def\K{\mbox{\mbf K}}
\def\E{\mbox{\mbf E}}
\def\F{\mbox{\mbf F}}

\def\x{\mbox{\mbf x}}
\def\y{\mbox{\mbf y}}
\def\z{\mbox{\mbf z}}

\def\p{\mbox{\mbf p}}

\def\zero{\mbox{\bf 0}}

\newcommand{\mbf}[1]{\mbox{\boldmath $#1$}}

\draft

\title{The effect of noise on parameter estimation in systems with 
complicated dynamics}

\author{Justin Goodwin}
\address{Department of Mathematics and Computer Science, University
of Puget Sound, Tacoma, WA 98414}

\author{Reggie Brown}
\address{Department of Physics and Department of Applied Science,
College of William and Mary, Williamsburg, VA 23187-8795}

\author{Lutz Junge}
\address{Drittes Physikalisches Institut, Universit\"{a}t
G\"{o}ttingen, B\"{u}rgerstrasse 42-44, D-37073 G\"{o}ttingen,
Germany}

\date{\today}
\maketitle

\begin{abstract}
Changes in parameters of a physical device can eventually give 
way to catastrophic failure.  This paper discusses a parameter
estimation method based on synchronization between a model and
time series data.  In particular, we examine the robustness of
the method to additive noise in the data.
\end{abstract}


\section{Introduction}
Scientists and engineers are often faced with the task of estimating 
the parameters of a physical device using time series data obtained
from the device, and a mathematical model for the dynamics of the 
device.  Because the dynamics of a model is governed by its parameters,
parameter estimation amounts to adjusting the parameters in the model
until its dynamics accurately mimics those of the device.  Then, the 
parameter values of the device are assumed to be identical to those
of the model.

Parameter estimation is motivated, in part, by the fact that
the dynamics of all physical devices change over time.  If the
change arises from normal ``wear and tear'' then it is typically
a slow change which eventually gives way to a rapid catastrophic
failure.  Parameter estimation aids in avoiding catastrophic
failure in two ways.  First, one can determine regions of parameter
space associated with failure by estimating and tracking parameter
values as devices are driven to failure.
Second, by estimating parameter values of a device at regular intervals
(and tracking these values) one can determine if and/or when 
the device approaches failure.

Currently, most devices in the field operate in regimes where 
their dynamics are regular (i.e. periodic or quasi-periodic).  For
these situations, slow changes in parameter values can be detected
by observing shifts in the frequencies and amplitudes of Fourier 
spectra from time series measurements.  Recently, there has been
speculation about devices whose normal operating regimes have
complicated and/or chaotic dynamics.  Examples of such devices
are bearings under high stress, electronic components with 
nonlinear response functions, mixing and electrochemical reactions,
and high speed cutting tools.
For these examples, complicated dynamics are unavoidable, and may
even be desirable.  More importantly, Fourier spectra will not
exhibit sharp peaks.  Therefore, other methods for estimating and
tracking changes in, parameter values are needed.

This paper discusses parameter estimation for systems whose underlying
dynamics are complicated and/or chaotic.  
An early example of parameter estimation in the nonlinear dynamics
literature is Hudson~{\em et.al.}~\cite{hkaklf}.  In this work the
authors constructed neural net models that mimicked a few of the period 
doubling bifurcations found in an electrochemical experiment.

Recently, Baker~{\em et.al.}~\cite{bgb} proposed a least squares 
approach that requires time series vectors which are simultaneous 
measurements of all of the phase space variables.  The approach also
requires numerical approximation of derivatives from the time series.
Horbelt~{\em et.al.}~\cite{htm} (building on work by Baake~{\em 
et.al.}~\cite{bbbb}) implemented a shooting method.  They used the
method, and experimentally obtained data, to estimate parameters
for a model of calcium release in muscle cells .  Also, a novel 
method using symbolic dynamics has been developed for systems that
exhibit complicated temporal or spatio-temporal dynamics~\cite{ttb}.

The approach discussed in this paper estimates the parameters of a
physical device by selecting values which yield ``the best 
synchronization'' between a time series from the device and a 
mathematical model for the device.  The method is related to Kalman
filtering~\cite{brogan} and relies on two ideas.  First, if two 
dynamical systems are identical then it is
possible for them to synchronize (i.e., both systems follow 
identical phase space trajectories).  Second, if this behavior 
is stable then (for the noise free case) deviations from 
synchronous behavior converge to zero as differences in parameter
values between the two systems converge to zero.  These ideas
imply that (in the absence of modeling errors) the best estimate
for the parameter values of a device are those of the model 
which best synchronizes to a time series from the device.

The synchronization approach to parameter estimation has been 
discussed by Parlitz~{\em et.al.}~\cite{parlitz,pjk} as well
as Maybhate and Amritkar~\cite{ma}.  The work discussed in this 
paper focuses on how noise in experimental data effects the accuracy
of parameter estimation.  Refs.~\cite{bgb,parlitz,pjk,ma} 
either did not discuss noise, or only examined small noise levels.

Furthermore, our implementation differs from the ones in previous
papers.  For example, Refs.~\cite{htm,pjk,ma} used various down
hill search methods to minimize the cost function.  However, it
is known~\cite{ttb,ershov} (and we have observed in the numerical
experiments discussed below) that a cost function can have many 
local minima where simple down hill methods can be trapped.  We
avoid this problem in some of our work by using annealing to
search the parameter space. 

Another important difference between our approach and 
Refs.~\cite{parlitz} and \cite{ma} concerns the time series used
in the estimation.  The previous authors assume the experimental
measurements are continuously recorded.  Thus, the time series has
measurements at every value of time.  When performing numerical
experiments with this type of time series it is reasonable to
replace the time series by a set of ODE's.

In our work we assume the time series consists of experimental 
measurements obtained at discrete increments of time (given by
a sampling interval).  When performing numerical experiments with
this type of time series one can not replace the time series by
a set of ODE's.  Rather, one must use an approach that explicitly
accounts for the discreteness of the time series.

Furthermore, Ref.~\cite{ma} assumes that one of the phase space
variables can be measured.  We only assume a scalar time series
corresponding to some experimental measurement.  It does {\em 
not} have to be one of the phase space variables.  Finally, our 
numerical experiments simultaneously estimate several parameters,
while Ref.~\cite{ma} typically estimates only one or two parameters
for a three dimensional model.  We have tested our methods on a 
high dimensional systems ( an 11~dimensional generalized
R\"{o}ssler model).

Although modeling errors are inevitable in any mathematical
description of a device, this paper does not address this
issues.  We conjecture that noise is the dominant obstacle 
associated with parameter estimation.  Our examples represent 
electronic circuits, chemical reactions, and a high dimensional
version of the R\"{o}ssler system.  The results indicate that
this approach to parameter estimation holds promise for real
applications.

The remainder of this paper is organized as follows.  In 
Section~\ref{sync} we define and discuss synchronization between
two or more dynamical systems.  In Section~\ref{estimate} we 
discuss parameter estimation via synchronization.  Examples
and the effects of noise are given in Section~\ref{examples}.
Conclusions are presented in Section~\ref{sum}.


\section{Synchronization}
\label{sync}
In this section we define and discuss synchronization between two
coupled deterministic dynamical systems.   The numerical experiments
use two different synchronization techniques, each of which is 
discussed.  The functional forms of
the uncoupled systems are identical, so they only differ in the
values of their respective parameters.  The coupling between the
systems is denoted by $\E(\x,\y)$, where $\x, \y \in \R^d$ are
the instantaneous states of the two systems.  The equations of
motion for this arrangement are
\begin{eqnarray}
\label{drive}
\frac{d \x}{dt} &=& \F(\x ; \p^*) \\
\label{response}
\frac{d \y}{dt} &=& \F(\y ; \p) + \E(\x,\y) .
\end{eqnarray}

Equation~(\ref{drive}), the drive system, represents the physical
device, while Eq.~(\ref{response}), the response system, represents
a mathematical model of the device.  The parameters of the device
and model are represented by $\p^*, \p \in \R^n$, respectively.
The coupling shown in Eqs.~(\ref{drive}) and (\ref{response}) is
called drive--response
coupling, and is natural for our application.  If the coupling in
Eq.~(\ref{response}) vanishes when $\x = \y$ (i.e., $\E(\x,\x) =
\zero$) then synchronization is defined as $\x(t_*) = \y(t_*)$ at
some time $-\infty < t_* < \infty $.  If this occurs then, in 
the absence of noise, determinism implies that $\x(t) = \y(t)$
for $t > t_*$.

However, the systems we examine are chaotic, no model of a 
physical device can ever be exact, the state of a model can never
exactly match the state of a device, and noise is never absent
from real data.  Therefore, synchronization between uncoupled 
chaotic systems never really exists because approximation 
synchronization ($\x \simeq \y$) between chaotic systems is 
always unstable.

These remarks imply that the coupling between Eqs.~(\ref{drive})
and (\ref{response}) is necessary to stabilize synchronous motion.
Over the last several years it has become increasingly clear 
that if $\p = \p^*$ then there are many choices of $\E(\x,\y)$
which will result in stable synchronization~\cite{chaos}.
For these types of coupling, if $\| \x - \y\|$ is small then
$\lim_{t \rightarrow \infty} \| \x - \y \| = 0,$
where $\| \bullet \|$ denotes the Euclidean norm.

One form of coupling that satisfies $\E(\x, \x) = \zero$ and  
often yields stable synchronization is
\begin{equation}
\label{coupling}
\E(\x,\y) = \K [ h(\x) - h(\y)] ,
\end{equation}
where \K\ (the so called gain vector) could have only one 
nonzero component.  Thus, the coupling between the drive and 
response systems could involve only one component of $\F(\y; \p)$.

It is usually impossible to simultaneously measure all of the
state variables of a physical device.  Typically, one 
experimentally measures a scalar function of the state of the
device.  To make contact with this more realistic case we
introduce the measurement functions, $h(\x)$ and $h(\y)$.  
These functions represent the experimental measurement process
which transforms the instantaneous state of the device or model
into a scalar measurement.

The time series, $h[\x(t_n)]$, represents scalar data measured 
at discrete times $t_n = n \Delta t$, $n=1,\,2, \ldots$, where 
$\Delta t$ is the sampling interval.  In contrast, numerical 
integration of the model requires $\E$ at arbitrary values of 
$t$.  To overcome this problem we have used two different methods
for calculating $\E$ at times not equal to one of the $t_n$'s.
One method selects all data within
a time window centered at $t$.  Least squares is used to fit the 
data points within this window to a polynomial, and the 
polynomial is used to approximate $h[\x(t)]$~\cite{polfit}.  For
this method the response system, Eq.~(\ref{response}), is 
continuously forced by the time series.

The second method is called sporadic driving~\cite{pksj}.  Here,
$\E(\x,\y) = \zero$ if $t$ is not equal to one of the $t_n$'s.
If $t$ is equal to one of the $t_n$'s then \E\ is given by 
Eq.~(\ref{coupling}).  For this method, the response system (the
model) is discontinuously kicked by the time series at multiples
of the sampling interval, and evolves freely between kicks.

Although the two methods differ in their implementation, each 
leads to stable synchronization.  Furthermore, each is a natural
choice when confronted with a drive system that is a time series
instead of ODE's.


\section{Parameter Estimation}
\label{estimate}
Assume we are given a time series, $h[\x(t_n)]$ for $n = 1,\,
2, \ldots, N$, representing experimentally measured data.  Our
task is to select values for the model parameters such that its
dynamics are the same as those of the device which generated 
the data.  If the parameter values of the device are $\p^*$,
and the estimated values are $\hat{\p}$, then the goal of 
parameter estimation is to achieve $\hat{\p} = \p^*$.

The synchronization method uses coupling that guarantees stable 
synchronization when $\p = \p^*$, and the drive system is an ODE.
This amounts to determining \K\ in Eq.~(\ref{coupling}).  To find
$\hat{\p}$ we choose parameters, \p, which minimize
\begin{equation}
\label{serror}
\chi (\p,\p^*) = \left[ \frac{1}{M} \sum_{n=1}^M (h[\x(n)] -
h[\y(n)])^2 \right]^{1/2} .
\end{equation}
Equation~(\ref{serror}) explicitly notes that deviations
between measured output from the model and measured output
from the devices are functions of the parameters of the model,
\p, and the device, $\p^*$.  This dependence arises from the 
dependence of the trajectories, \x\ and \y, on the parameters
in Eqs.~(\ref{drive}) and (\ref{response}).  We denote the
value of \p\ which minimizes Eq.~(\ref{serror}) by $\hat{\p}$,
and note that $\chi(\p^*, \p^*) = 0$ in the absence of noise.

The numerical experiments typically use $M/N \sim 1/3$ or 2/3.
The remaining $N-M$ points are treated as a transient phase 
used to initialize synchronization.

The first two examples used annealing to search parameter space
and avoid becoming trapped in local minimum far from 
$\p^*$~\cite{nr}.  If $\hat{\p}$ is from such a local minimum 
then we expect the dynamics of the model to be very different 
from that of the device.  To detect this, we performed a simple 
post-minimization test.  After finding $\hat{\p}$ we generated
a time series, $h(\y)$, from the model and used time delay 
embedding of $h(\x)$ and $h(\y)$ to reconstruct attractors
for the data and the model~\cite{say,abst}.

To test the similarity of the attractors, we selected a point
on the attractor reconstructed from the model time series,
and calculated the distance between it and its 
nearest neighbor in the attractor reconstructed from the 
experimental data.  If the dynamics of the model and the
device are the same then every point on the attractor 
reconstructed from the model time series will have a near 
neighbor on the attractor reconstructed from the experimental
data.  More importantly, the distance to the nearest neighbor
will not be large.

However, if the dynamics of the model and the device are 
different then some points on the model attractor will be 
far away from {\em all} points on the experimental attractor,
and the distance to the nearest neighbor will be large.  If
this is the case then we choose a new initial condition in
parameter space and minimized Eq.~(\ref{serror}) again.

The last example used a gradient descent method (POWELL in
Ref.~\cite{nr}) to search the parameter space~\cite{pjk}.  
To avoid a local minima, we change each parameter by an
amount of order $\sim 10^{-3}$, and the descent is started
again.  This procedure is repeated until we find a value 
for $\hat{\p}$ that is stable against such perturbations.

Numerical experiments indicate that we typically find a value
of $\hat{\p}$ that is near $\p^*$ using only one or two 
initial conditions in parameter space.  (The annealing 
procedure we use is probably not the best implementation 
of annealing.  We believe that a more sophisticated 
implementation would remove this problem~\cite{ingber}.) 
For the passivation example discussed below, the gradient
descent method frequently fails to find the global minimum
on the first try.

The numerical experiments use the following coupling between
the drive and response systems, 
\begin{equation}
\label{gain}
\E(\x,\y) = \K [ (h(\x) + \eta) - h(\y)] .  
\end{equation}
Here, $\eta$ represents noise, and $(h(\x) + \eta)$ represents
an experimental measurement that has been contaminated with 
additive noise.  In the numerical experiments $\eta = \sigma 
\Sigma N(0,1)$, where $N(0,1)$ denotes random numbers normally 
distributed with mean zero and unit standard deviation.  $\Sigma$
represents the size of the experimental signal, and was equated
to the standard deviation of the clean time series, $h(\x)$. 

We could have examined dynamic noise.  However, dynamic noise
is equivalent to modeling errors which change instantaneously
in time.  Since we have chosen not to address modeling errors
it is reasonable to ignore this type of noise.


\begin{figure}
\begin{center}
\leavevmode
\psfig{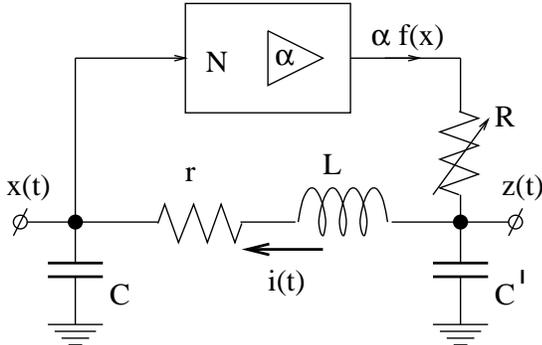}
\end{center}
\caption{A block diagram for an electronic circuit modeled by
Eqs.~(\protect{\ref{circ}}) and (\protect{\ref{def_F}}).
\label{fig.block}}
\end{figure}

\section{Examples}
\label{examples}
We now discuss the effect of noise on the accuracy of parameter
estimation.  The first example used a model of the electronic 
circuit shown
in Fig.~\ref{fig.block}.  This circuit consist of a nonlinear 
amplifier, $N$, which transforms input voltage, $x$, into
output, $\alpha f(x)$~\cite{rulkov}.  The parameter, $\alpha$,
characterizes the gain around $x=0$.  The amplifier has linear
feedback consisting of a series connection to a low-pass filter,
$RC'$, and a resonant circuit $LC$.  The system has been 
studied experimentally, is known to exhibit chaotic dynamics,
and has been shown to synchronize with a copy of itself for 
many coupling schemes~\cite{rulkov}.

It is known that a good model for this circuit is the following
system of ODE's
\begin{eqnarray}
\label{circ}
\frac{dx_1}{dt} &=& x_2 \nonumber \\
\frac{dx_2}{dt} &=& -x_1 - \delta x_2 + x_3 \\
\frac{dx_2}{dt} &=& \gamma [F(x_1) - x_3] - \beta x_2
\nonumber
\end{eqnarray}
with $F(x)$ defined by
\begin{equation}
\label{def_F}
F(x) = \left\{ 
\begin{array}{ll} 
0.528 \alpha &  x < -1.2, \\ 
\alpha x (1 - x^2) & -1.2 < x < 1.2,  \\
-0.528 \alpha & x > 1.2
\end{array} \right.
\end{equation}

\begin{figure}
\vspace{0.0in}
\begin{center}
\leavevmode
\psfig{figure=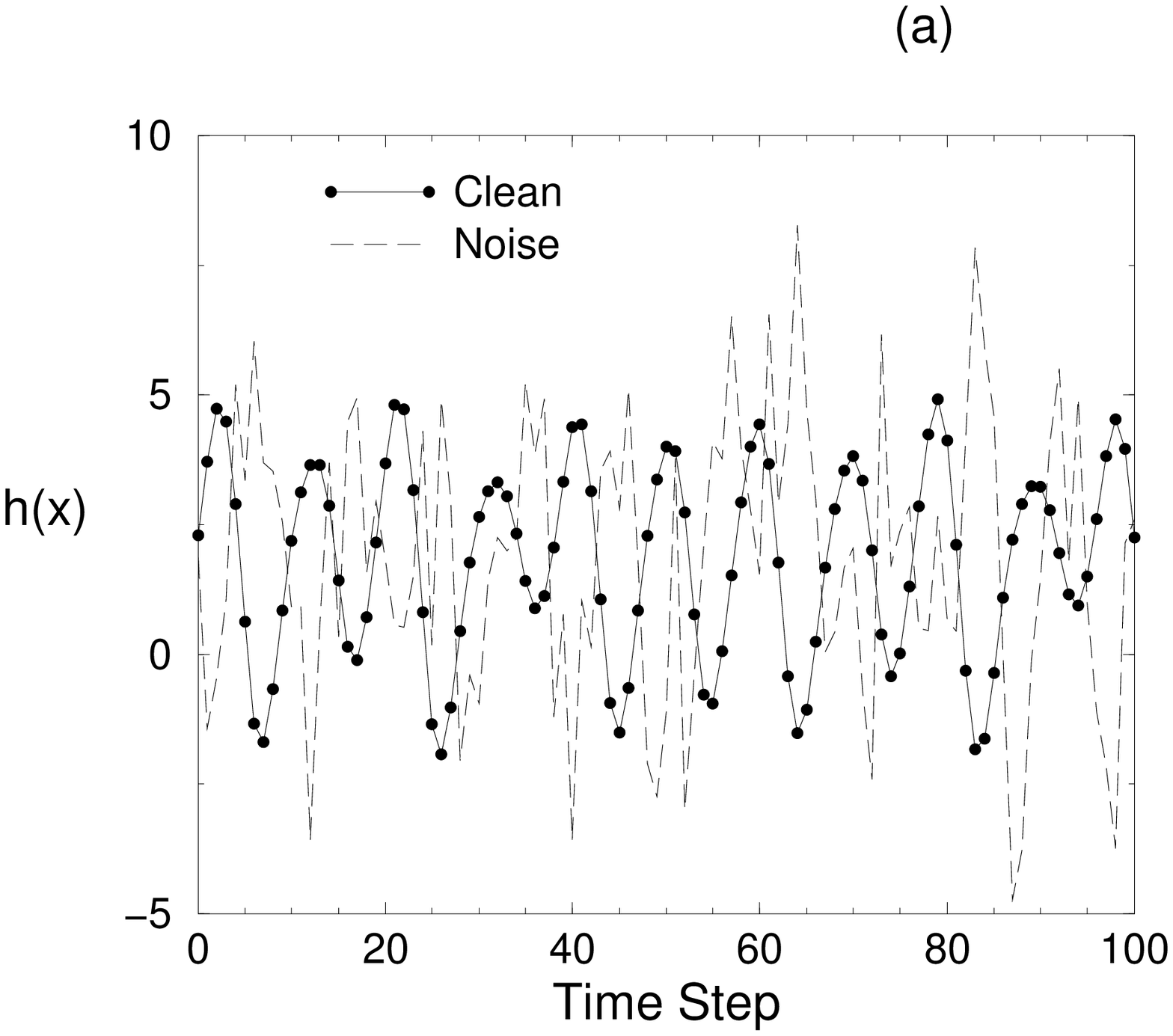,width=3.0in}
\psfig{figure=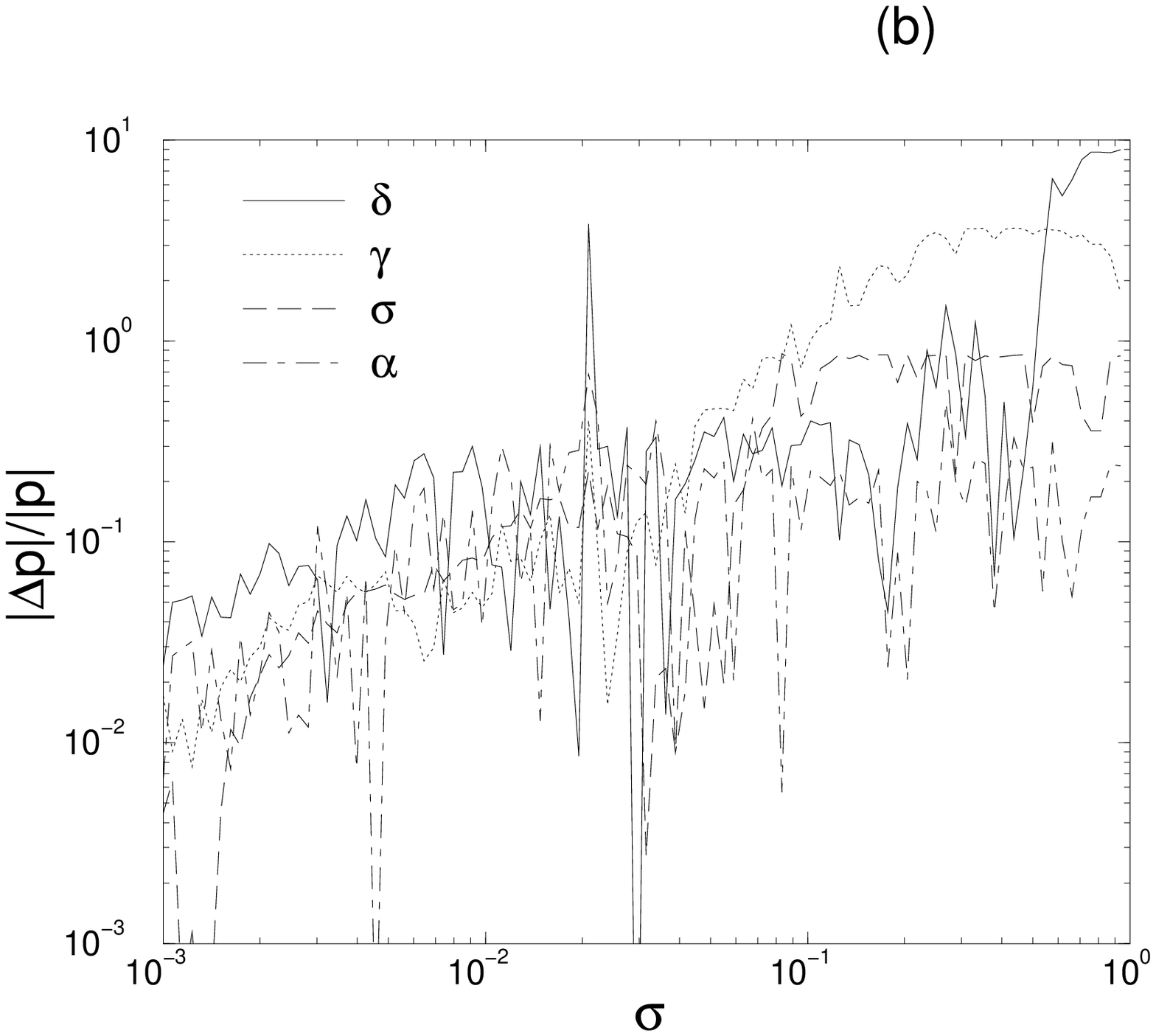,width=3.0in}
\end{center}
\caption{The electronic circuit model.  (a) Short samples of the
time series used in the numerical experiments.  The noisy time 
series represents 0~dB noise. (b) The relationship between the 
relative error in the parameter estimates and the size of the noise.
Here, $\Delta \p = \hat{\p} - \p^*$.
\label{fig.cir}}
\end{figure}

The numerical experiments use Eqs.~(\ref{circ}) and 
(\ref{def_F}) for the response system, continuous coupling
given by Eq.~(\ref{gain}), and parameter values $\p^* = [
\delta,\, \gamma,\, \beta,\, \alpha]$ with
$\delta = 0.43$, $\gamma = 0.1$, $\beta = 0.72$, and
$\alpha = 16$.  For this circuit it is straightforward to 
measure the voltages associated with \x.  However, to make
contact with other experimental devices we used the 
following (arbitrarily selected) measurement function
\begin{eqnarray}
\label{def_h_cir}
h(\x) = 2 x_1 + 3 x_2 + 0.5 x_3 .
\end{eqnarray}
The gain vector is $\K = [0, 0, 10]$.  In the absence of noise
this coupling leads to stable synchronization when the drive
{\em and} response systems are ODE's.  The sampling interval is
$\Delta t = 0.5$ and the data is shown in Fig.~\ref{fig.cir}(a).
For this example, $\sigma$, the size noise ranged from a low of
0.001 (representative of very clean data) to a high of 1 
(representative of very noisy data).  

We use an Adams-Bashforth-Moulton code numerically integrate the
model~\cite{abm}.  The initial guess for \p\ in the annealing 
procedure is selected at random from within the range $p_1 \in
[0, 1]$, $p_2 \in [0,2]$, $p_3 \in [0,5]$ and $p_4 \in [0,20]$.
In the absence of noise, the method correctly estimated $\hat{\p}
= \p^*$ to within machine precision.  The numerical experiments 
with noise examined 100 different value of $\sigma$.  For each 
value we calculate $\hat{\p}$.  The results of the numerical 
experiments (see Fig.~\ref{fig.cir}(b)) indicate that $\|\hat{\p}
- \p^* \| \propto \sigma \| \p^* \|$ over much of the range of 
noise strengths.  This result confirms earlier 
theoretical predictions~\cite{brt2}).  

The next example uses a model for metal passivation.  Passivation
is a loss of chemical reactivity associated with metal
corrosion in aqueous solutions.  It is an electrochemical reaction
that exhibits the full zoo of nonlinear dynamical behavior, including
periodic cycles, period doubling, multistability, hysteresis, chaos.
These behaviors are not unusual in electrochemical 
reactions~\cite{to,hlps,koper}.  

\begin{figure}
\vspace{0.0in}
\begin{center}
\leavevmode
\psfig{figure=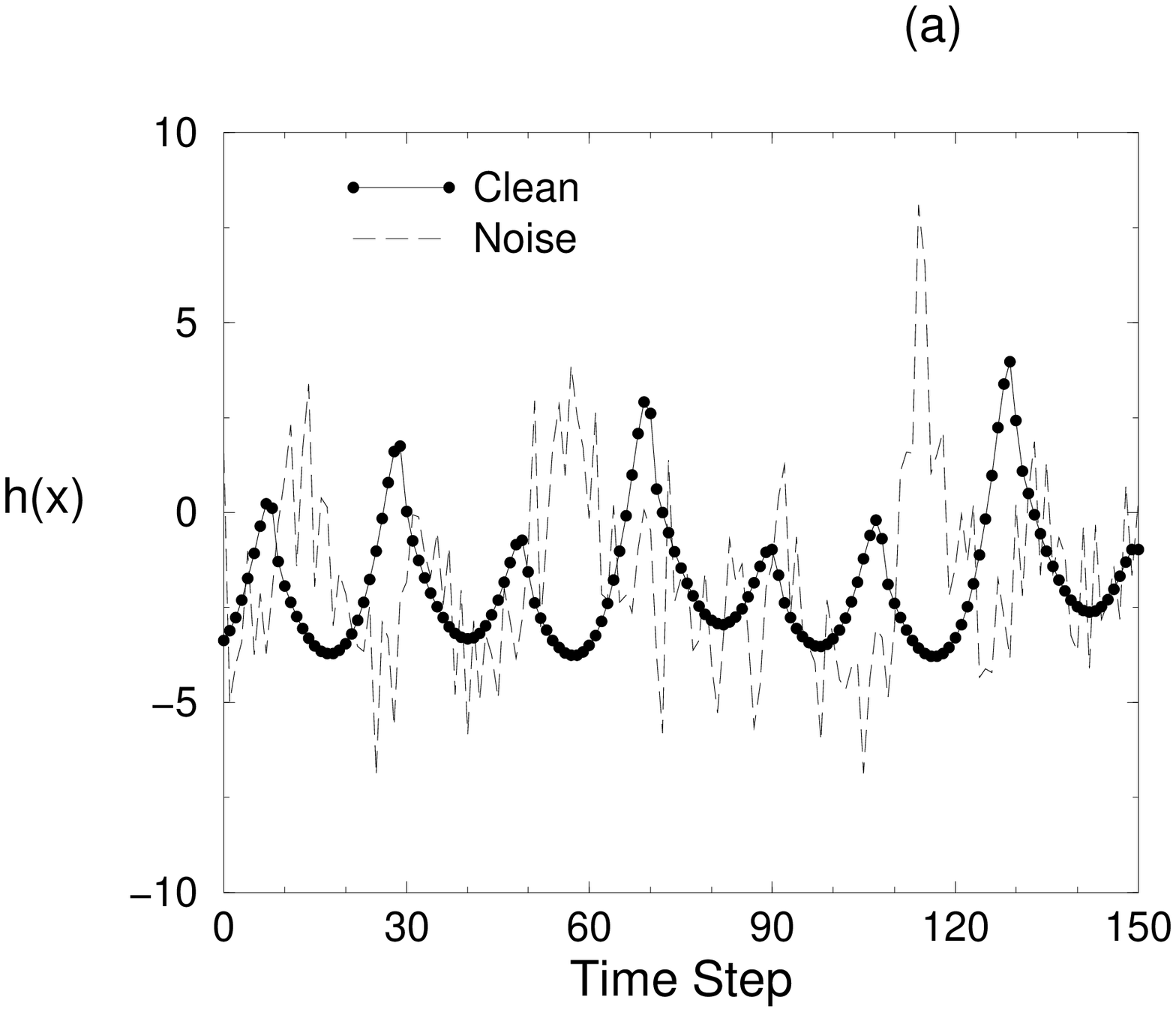,width=3.0in}
\psfig{figure=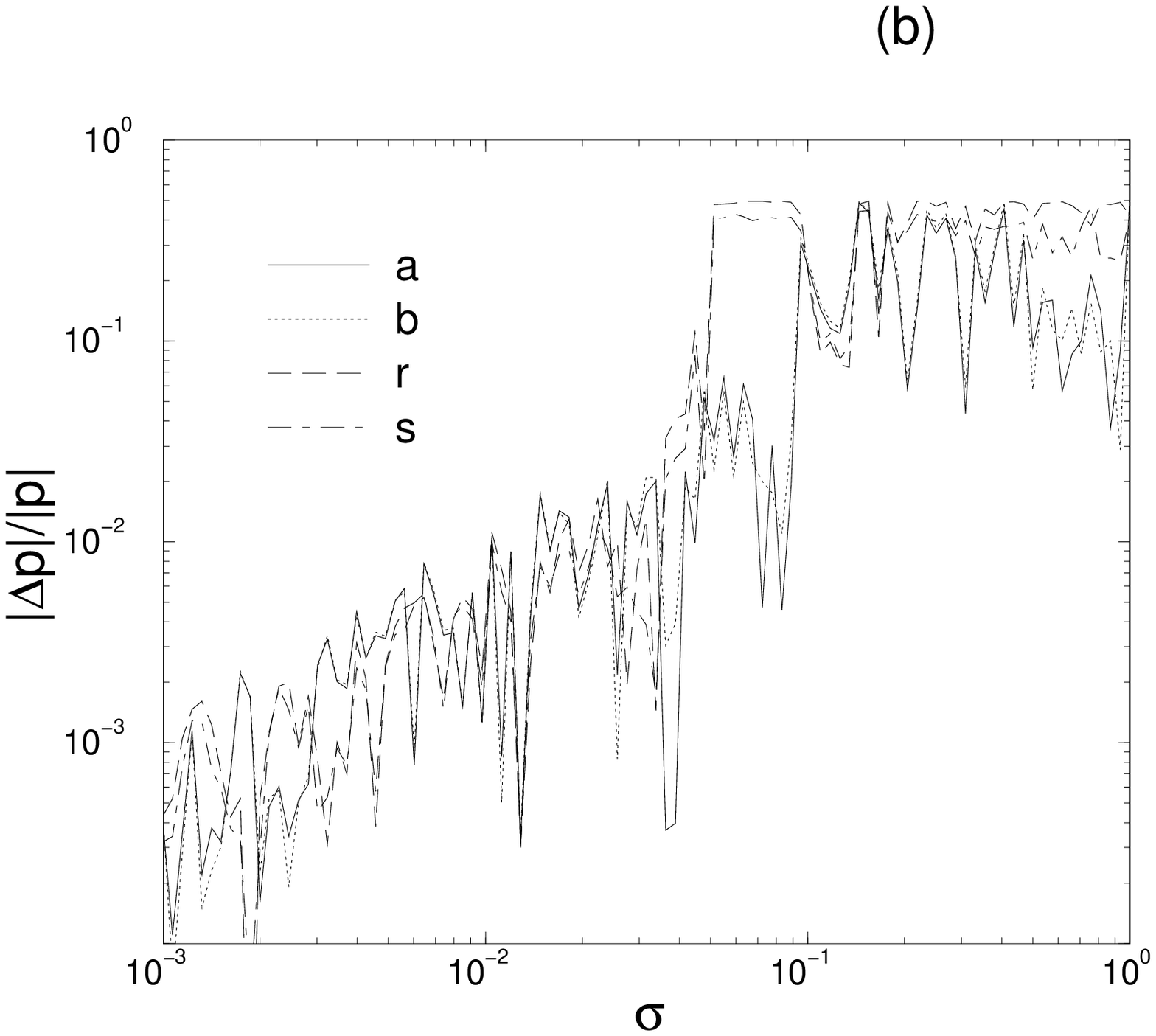,width=3.0in}
\end{center}
\caption{The metal passivation model.  (a) Short samples of the
time series used in the numerical experiments.  The noisy time 
series represents 0~dB noise. (b) The relationship between the 
relative error in the parameter estimates and the size of the noise.
Here, $\Delta \p = \hat{\p} - \p^*$.
\label{fig.pass}}
\end{figure}

We use the following passivation model~\cite{mccoy}
\begin{eqnarray}
\frac{d Y}{dt} & = & a \theta_{\rm M} - b Y \nonumber \\
\frac{d \theta_{\rm OH}}{dt} & = & Y \theta_{\rm M} - [r + \exp
(-\beta \theta_{\rm OH})] \theta_{\rm OH} + 2 s \theta_{\rm O}  
\theta_{\rm M} \\
\frac{d \theta_{\rm O}}{dt} & = & r \theta_{\rm OH} - s \theta_{\rm
O} \theta_{\rm M} , \nonumber 
\end{eqnarray}
where $\theta_{\rm M} \equiv (1 - \theta_{\rm O} - \theta_{\rm 
OH})$, and the parameters $\p^* = [a,\, b,\, r,\, s]$ are $a = 2 
\times 10^{-4}$, $b = 10^{-3}$, $r = 2 \times 10^{-5}$, and $s = 
9.7 \times 10^{-5}$.  We fixed the last parameter at $\beta = 5$.
This example is subtle because of the small size of the parameters
and the attractor.  To overcome this we rescale the variables by
subtracting the mean and dividing by a constant roughly equal the 
width of the attractor in that coordinate.  Time is also rescaled 
by a factor of 10,000~\cite{scale}.  The coupling between drive and
response system uses
\begin{eqnarray}
\label{def_h_chem}
h(\x) = 0.2 x_1 + x_2 + 0.5 x_3.
\end{eqnarray}
with gain vector $\K = [0, 100, 0]$.  Once again $h(\x)$ is an
arbitrarily chosen function, and in the absence of noise this
coupling leads to stable synchronization when the drive {\em 
and} response systems are ODE's.  The sampling interval of the 
numerical experiments is $\Delta t = 0.01$ (in rescaled time) 
and the data is shown in Fig.~\ref{fig.pass}(a).

For this example, $\sigma$, the size noise ranged from a low of
0.001 (representative of very clean data) to a high of 1 
(representative of very noisy data).  
100 different value of $\sigma$ were examine.  The results,
shown in Fig.~\ref{fig.pass}~(b), indicate that the relative
accuracy of parameter estimation is two digits or more up to
$\sigma \sim 0.1$ and about 30~\% for 0~dB noise.  This
surprisingly robust result implies that synchronization is worthy
of further investigation as a parameter estimation technique.

\begin{figure}]
\vspace{0.0in}
\begin{center}
\leavevmode
\psfig{figure=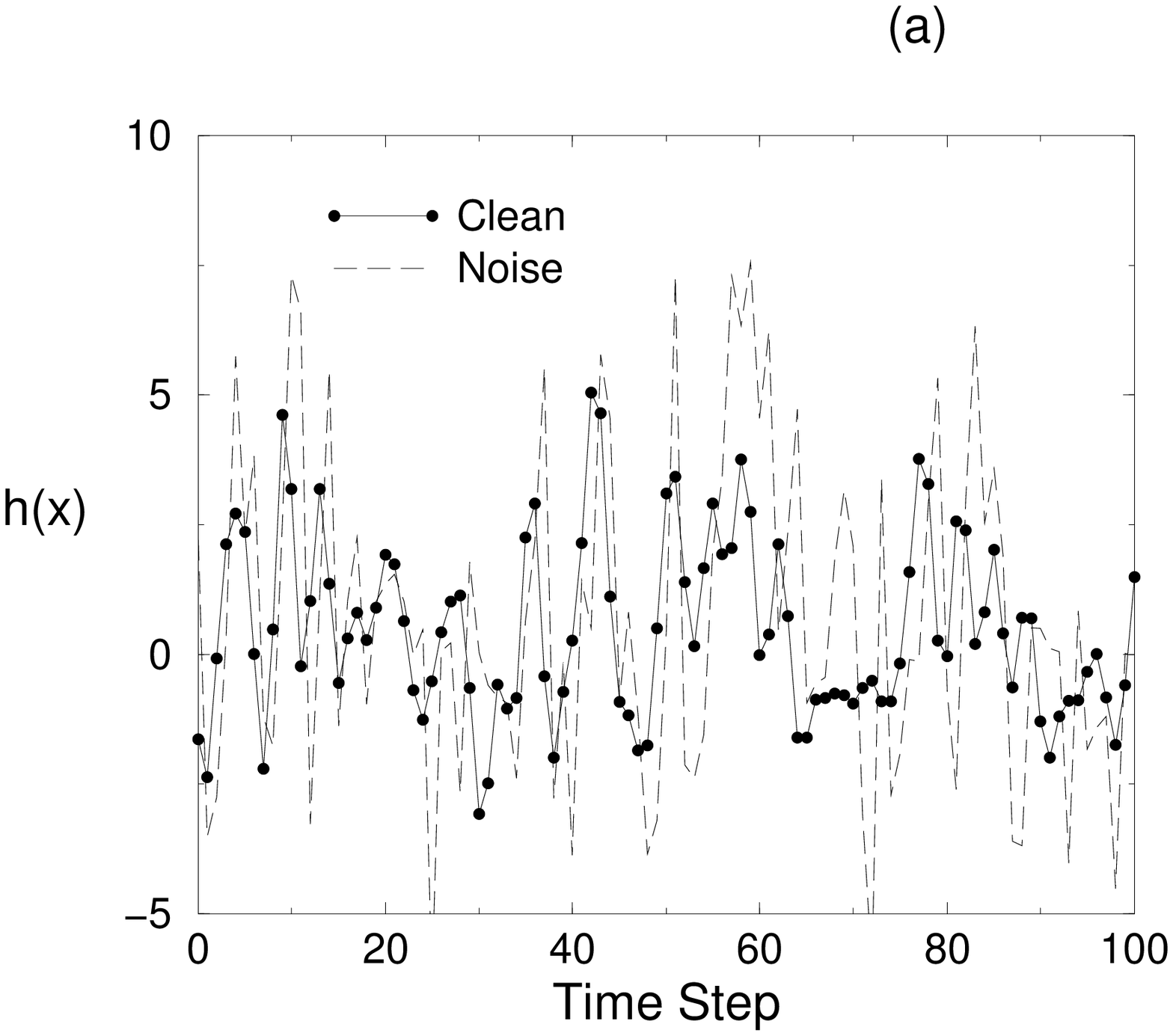,width=3.0in}
\psfig{figure=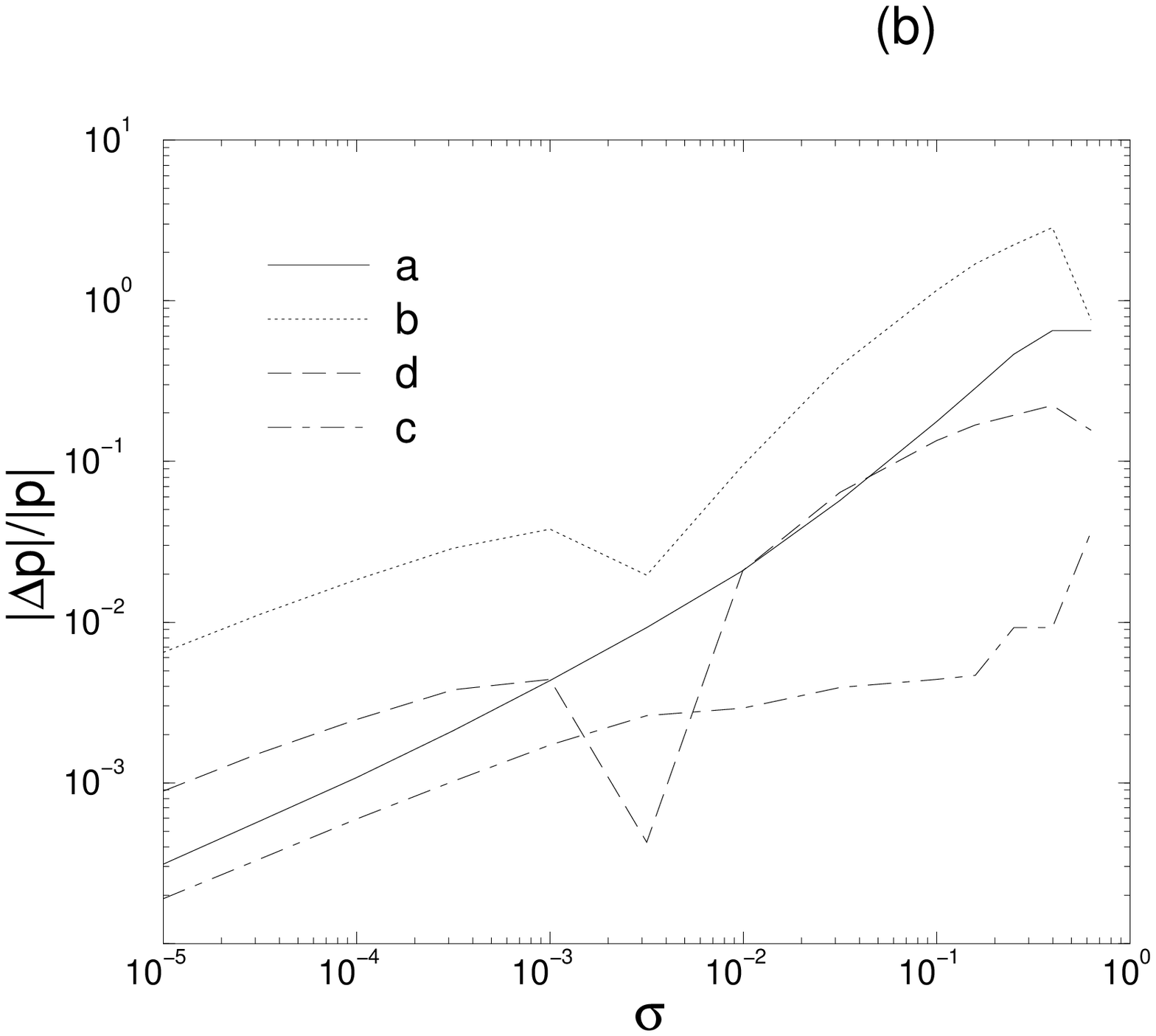,width=3.0in}
\end{center}
\caption{The hyperchaotic generalized R\"{o}ssler model.  (a) Short samples of the
time series used in the numerical experiments.  The noisy time 
series represents 0~dB noise. (b) The relationship between the 
relative error in the parameter estimates and the size of the noise.
Here, $\Delta \p = \hat{\p} - \p^*$.
\label{fig.ross}}
\end{figure}

The final example is a high dimensional generalization of the
R\"{o}ssler system \cite{baier}
\begin{eqnarray*}
\frac{dx_1}{dt} & = & - x_2 + a \, x_1 \\
\frac{dx_j}{dt} & = &  x_{j-1} - x_{j+1} \:\:\:\:\:\:\:\: (j=2,\,
\ldots, L-1) \\
\frac{dx_L}{dt} & = & b + d \, x_L (x_{L-1} - c) .
\end{eqnarray*}
Here, $\p^* = [a,\, b,\, c,\, d]$ with $a = 0.29$, $b = 0.1$, $c = 
2$, and $d = 4$.   This system is
the well known R\"{o}ssler system with an additional $L-3$~degrees of
freedom.  It has a stable chaotic attractor when $L$ is odd.  Our
numerical experiments used $L=11$, yielding a hyperchaotic attractor with a
Lyapunovdimension of $D_L\approx 10.1$.  For this example we used for synchronization
the method of sporadic driving with $h(\x)= x_1$ as the drive signal and
a sampling interval of $\Delta t = 1$.
The results were calculated for 15 different noise levels ranging
from $\sigma = 10^{-5}$ to $\sigma = 0.63$.  The
results are shown in Fig.~\ref{fig.ross}. As in the other examples, we observe a linear
scaling of the parameter relative error $\frac{\Delta \p}{\p}$ with the noise 
strength $\sigma$ and find good estimates up to $\sigma\sim 0.15$ 
The quality of the parameter estimates in relation to the noise strength is almost the 
same as for the electronic circuit example. This result confirms that the method
can be applied to highdimensional systems, even if only very noisy time series data is
available.


\section{Conclusion}
\label{sum}
In this paper we have examined the effects of noise on the accuracy 
of parameter estimation for physical devices.  The parameter estimation
method we examined uses experimentally measured time series and a 
mathematical model of the device.  The estimates returned by the
method are those which yields the smallest deviation from 
synchronization between the dynamics of the model and the 
measured time series. 

Numerical experiments indicate that this method is robust to
additive noise in the time series.  We find that even at 0~dB 
signal to noise ratio we can still obtain reasonable accuracy in
the parameter estimates. The method works also for noisy time series 
coming from highdimensional systems.
The obvious next step is to test this
method ``at sea'', by using experimental data coming from electronic~\cite{pjk}, 
physical, chemical, \dots systems.


\begin{references}
\bibitem{polfit} This code, polfit.f, can be found in SLATEC, a public 
domain source code library.
\bibitem{abm} This code, deabm.f, can be found in SLATEC, a public 
domain source code library.
\end{references}
\end{document}